\documentclass[journal]{vgtc}                

\ifpdf
  \pdfoutput=1\relax                   
  \usepackage{graphicx}                
  \DeclareGraphicsExtensions{.pdf,.png,.jpg,.jpeg} 
\else
  \usepackage{graphicx}                
  \DeclareGraphicsExtensions{.eps}     
\fi%

\graphicspath{{figures/}{pictures/}{images/}{./}} 

\usepackage{microtype}                 
\PassOptionsToPackage{warn}{textcomp}  
\usepackage{microtype}                 
\usepackage{textcomp}                  
\usepackage{mathptmx}                  
\usepackage{times}                     
\usepackage{fontawesome}
 
\usepackage{balance}  
\usepackage{cite}                      
\usepackage{tabu}                      
\usepackage{booktabs}
\usepackage{color,soul}
\usepackage[caption=false]{subfig}
\usepackage{float}
\usepackage{graphicx,wrapfig,lipsum}
\usepackage{tabu}                      
\usepackage{booktabs}
\usepackage{xcolor,soul}
\usepackage[caption=false]{subfig}

\usepackage{booktabs} 

\usepackage{multirow}
\usepackage{xspace}
\usepackage{colortbl}
\usepackage{bm}
\usepackage{amssymb}
\usepackage{pifont}
\newcommand{\cmark}{\ding{51}}%
\usepackage{microtype} 
\usepackage{mathptmx}
\usepackage{graphicx}
\usepackage{times}
\usepackage{enumitem}

\usepackage{xspace}
\setlist{nosep,topsep=4pt}

\newcommand{\MVS}{\emph{Manual View Specification}\xspace}

\newcommand{\demo}{\emph{Visualization by Demonstration}\xspace}
\newcommand{\tool}{\emph{Liger}\xspace}

\newcommand{\opAxis}{$\bm{O_{axis}}$\xspace}
\newcommand{\opMark}{$\bm{O_{mark}}$\xspace}
\newcommand{\opSwitch}{$\bm{O_{switch}}$\xspace}
\newcommand{\opFilter}{$\bm{O_{filter}}$\xspace}
\newcommand{\opSort}{$\bm{O_{sort}}$\xspace}

\newcommand{\ourParskip}{\vspace{0.33\baselineskip}}

\newcommand*{\glyph}[1]{
	\hspace{-.07cm}\raisebox{-0pt}{\includegraphics[height=6pt]{pictures/#1}}\xspace
}

\newcommand{\inlinebold}[1]{\ourParskip\noindent\textbf{\textit{#1.}}}

\onlineid{0}

\vgtccategory{Research}
\vgtcpapertype{please specify}

\title{Liger: Combining Interaction Paradigms for Visual Analysis}

\author{Bahador Saket, Lei Jiang, Charles Perin, and Alex Endert}
\authorfooter{
\item
Bahador Saket, Lei Jiang, and Alex Endert are with Georgia Tech. E-mail:\{saket,ljiang87,endert\} @gatech.edu.
\item
Charles Perin is with University of Victoria, E-mail: cperin@uvic.ca.
}

\shortauthortitle{Biv \MakeLowercase{\textit{et al.}}: Global Illumination for Fun and Profit}

\abstract{
Visualization tools usually leverage a single interaction paradigm (e.g., manual view specification, visualization by demonstration, etc.), which fosters the process of visualization construction. A large body of work has investigated the effectiveness of individual interaction paradigms, building an understanding of advantages and disadvantages of each in isolation. However, how can we leverage the benefits of multiple interaction paradigms by combining them into a single tool? We currently lack a holistic view of how interaction paradigms that use the same input modality (e.g., mouse) can be combined into a single tool and how people use such tools. To investigate opportunities and challenges in combining paradigms, we first created a multi-paradigm prototype (\tool) that combines two mouse-based interaction paradigms (manual view specification and visualization by demonstration) in a unified tool. We then conducted an exploratory study with \tool, providing initial evidence that people 1) use both paradigms interchangeably, 2) seamlessly switch between paradigms based on the operation at hand, and 3) choose to successfully complete a single operation using a combination of both paradigms. 

} 


\CCScatlist{ 
 \CCScat{K.6.1}{Management of Computing and Information Systems}%
{Project and People Management}{Life Cycle};
 \CCScat{K.7.m}{The Computing Profession}{Miscellaneous}{Ethics}
}

\vgtcinsertpkg

\begin{document}


\firstsection{Introduction}

\maketitle

In this paper, we investigate the challenge of combining interaction paradigms in desktop-based visualization tools with mouse input. 
Interaction is an essential part of visualizations tools, as it enables people to both construct visualizations and conduct data analyses~\cite{yi2007toward, dix1988starting}. There exists a wide variety of visualization tools, and these tools leverage a variety of interaction paradigms. 
In this paper, we use the term \textit{interaction paradigm} to refer to the process of how visualization construction is fostered in a tool. 
Although other terms such as ``interaction style'' or ``interaction model'' could also be used, we choose to use the term interaction paradigm similar to~\cite{saket2018evaluation,saketVbD}.


The visualization community has investigated the effectiveness of interaction paradigms implemented in visualization tools (e.g.,~\cite{grammel2013survey, Mendez:2018, Mendez:2017}). 
These studies have built an understanding of advantages and disadvantages of different interaction paradigms. 
However, studying these paradigms in isolation does not provide a holistic understanding of how one might use several paradigms together.
We do not know \textit{if} combining multiple interaction paradigms in a visualization tool is useful; nor \textit{how} multiple  paradigms can be used in combination. 
As a first step toward answering these questions, we investigate the use of two mouse-based interaction paradigms in visualization. 

The first paradigm we consider is \textbf{\MVS} \textbf{(MVS)}~\cite{saket2018evaluation}. 
MVS is arguably the most widely used paradigm, implemented in well-known visualization tools such as Tableau~\cite{Tableau} and Spotfire~\cite{ahlberg1996spotfire}. 
MVS enables people to manually specify mappings (from data to visual encodings) through GUI operations on collections of visual properties and attributes that are presented on control panels. 
For instance, to create a bar chart, one would specify the visualization type, then map data attributes onto axes, and map additional data attributes to visual encodings.
Many tools that have differences in user interface design~\cite{grammel2013survey} have a consistent underlying MVS interaction paradigm. For example, Tableau~\cite{Tableau} and Polestar~\cite{PoleStar} implement MVS by letting people drag and drop attributes onto shelves to set specifications; other tools such as Spotfire~\cite{ahlberg1996spotfire} implement MVS by letting people use dropdown menus. 
MVS tools let the person using the tool specify visual properties, and the system responds by generating the resulting view. 

The second paradigm we consider is \textbf{\demo} \textbf{(VbD)}~\cite{saket2018evaluation, saket_CGA, saketVbD}. 
With VbD, instead of specifying mappings between data attributes and visual encodings directly, people provide partial demonstrations of their intent to the visual representation using direct manipulation of visual marks.
For example, using VbD a person might convey their interest in mapping a data attribute to color by coloring one or more data points. 
The system then interprets the intent of the person and suggests or applies changes to the visualization. 
VbD leverages research showing the benefits of letting people create spatial representations of data points manually,
without formalizing the mappings between data and spatial constructs~\cite{huron2014constructive,shipman1999formality}.

Both MVS and VbD enable people to iteratively build visualizations by performing visualization operations.  
Such operations include switching between visualization techniques and mapping data attributes to visual encodings (e.g., size). 
However, MVS and VbD have intrinsic differences, and both have their own advantages and disadvantages~\cite{saket2018evaluation}.  
MVS tools are easy to learn and fast, because they have high external consistency~\cite{Grudin:1989,saket2018evaluation}. 
On the other hands, VbD tools have higher interaction expressivity, thus increase the levels of perceived control and engagement for the user~\cite{saket2018evaluation}. 
This leads us to consider how complementary these two paradigms are, and if it is possible to leverage the advantages of each, while limiting their respective disadvantages. 
However, we do not know \textit{how to offer the benefits of two different interaction paradigms in a unified visualization tool.} 


Little work has explored how multiple interaction paradigms that use \textit{the same input modality} (mouse for MVS and VbD) can be blended into a single visualization tool. 
We hypothesize that the expressivity provided by VbD can be a beneficial addition to tools that rely on the well-known MVS paradigm. 
Exploring such multi-paradigm tools poses several challenges.
First, it requires careful design and implementation considerations to ensure usability and proper combination of the two paradigms.
Second, it requires studying empirically the extent to which such interfaces facilitate common visual analysis tasks, as well as whether they lead to an improved user experience. 

To address these research questions, we first create \tool, a visual data exploration prototype that unifies the MVS and VbD paradigms. 
We use \tool as a testbed to investigate opportunities and challenges in combining multiple paradigms. 
Through the design and implementation of \tool, we exemplify how interaction paradigms can be blended to generate context that complements the individual paradigms. 
We then report a qualitative study of \tool with 10 participants that shows how people use both interaction paradigms for data exploration. Further, we discuss varying preferences for interaction paradigms, opportunities and challenges in multi-paradigm interfaces. 





\section{Related Work}

Several visualization process models explain the steps users follow to construct visualizations and conduct visual data analysis~\cite{card1999readings, Chi:interactionmodel, Carpendale:thesis}. 
The ``visualization reference model''~\cite{card1999readings} introduces the steps of \textbf{Raw Data Transformation}, \textbf{Data Table Transformation}, \textbf{Visual Properties Specification}, and \textbf{View Rendering} for creating and interacting with visualizations. 
This model and its variations (e.g.,~\cite{Chi:interactionmodel,Carpendale:thesis})
place \textit{Visual Properties Specification} before \textit{View Rendering}.
As a result, interaction paradigms following this approach ask users to first map data attributes to visual properties, then have systems render the views based on these mappings.
We use this observation to explain the differences between MVS and VbD in the next subsections (see Figure~\ref{fig:InteractionModel}).

\subsection{Manual View Specification (MVS)}
Following the visualization reference model, the MVS paradigm asks users to map data attributes to visual properties prior to rendering the view. This is the case with tools like MS Excel, Spotfire~\cite{ahlberg1996spotfire}, Tableau~\cite{Tableau} and Polaris~\cite{polaris:infovis00}. 
For instance, to create a scatterplot, users must specify the point visualization technique and map data attributes to the $x$ and $y$ axes. The system then generates the corresponding scatterplot. 

Grammel et al.~\cite{grammel2013survey} surveyed desktop-based visualization tools with mouse/keyboard input, from which
they extracted six categories of user interface design. 
Four of these categories are most relevant to our work: \textbf{template editors}, \textbf{shelf configuration}, \textbf{visual builder}, and \textbf{visual data flow}. 
Below, we describe these four designs, emphasizing that they are all variations of the MVS paradigm.

With the \textbf{template editor} design, \textit{``the user selects some data and then picks a predefined visual structure in which to represent it. The distinguishing criteria of this approach are the separation between the initial visualization selection steps and the refinement of the selected visualization''}~\cite{grammel2013survey}. 
This design requires users to specify data attributes before selecting a predefined visual structure. 
Tools like Many Eyes~\cite{viegas2007manyeyes} and MS Excel implement the template editor design. 

The \textbf{visual builder} design often consists of an empty canvas on which visual elements from a palette can be assembled -- similar to graphics editor tools such as Sketch and Adobe Illustrator.
Visual builder tools, like template editor tools, require users to specify visualization properties prior to rendering the final view. But the approach is different. 
First, users can draw a customized visual glyph and assemble visual elements together on the canvas. 
Then, they can manually bind graphical properties of the visual glyph to data attributes. 
Visualization authoring tools such as Data Illustrator~\cite{Liu:2018}, Data-Driven Guides~\cite{kim2017data}, and Charticulator~\cite{Charticulator} implement this design.

The \textbf{data flow} design lets users construct visualizations by connecting visual components through links. 
The final graph of connected components represents the dataflow and the final visual output. While the data flow design has a long history, recent visualization tools such as iVoLVER~\cite{Mendez:2017} have revisited this design.

The \textbf{shelf configuration} design lets users specify visual mappings on collections of visual properties and data attributes that are presented on control panels. 
The shelf configuration design can have variations.
For example, tools such as Tableau~\cite{Tableau} and Polestar~\cite{PoleStar} let people drag and drop attributes onto shelves to set specifications. Other tools such as Spotfire~\cite{ahlberg1996spotfire} do so by providing dropdown menus. 
Regardless of the implementation though, the users’ responsibilities are still to specify visual properties through graphical widgets presented on the interface.

\subsection{Visualization by Demonstration (VbD)}
\begin{figure}[!t]

\begin{minipage}[t]{1.0\linewidth}
	\centering
	\includegraphics[width=1.0\linewidth]{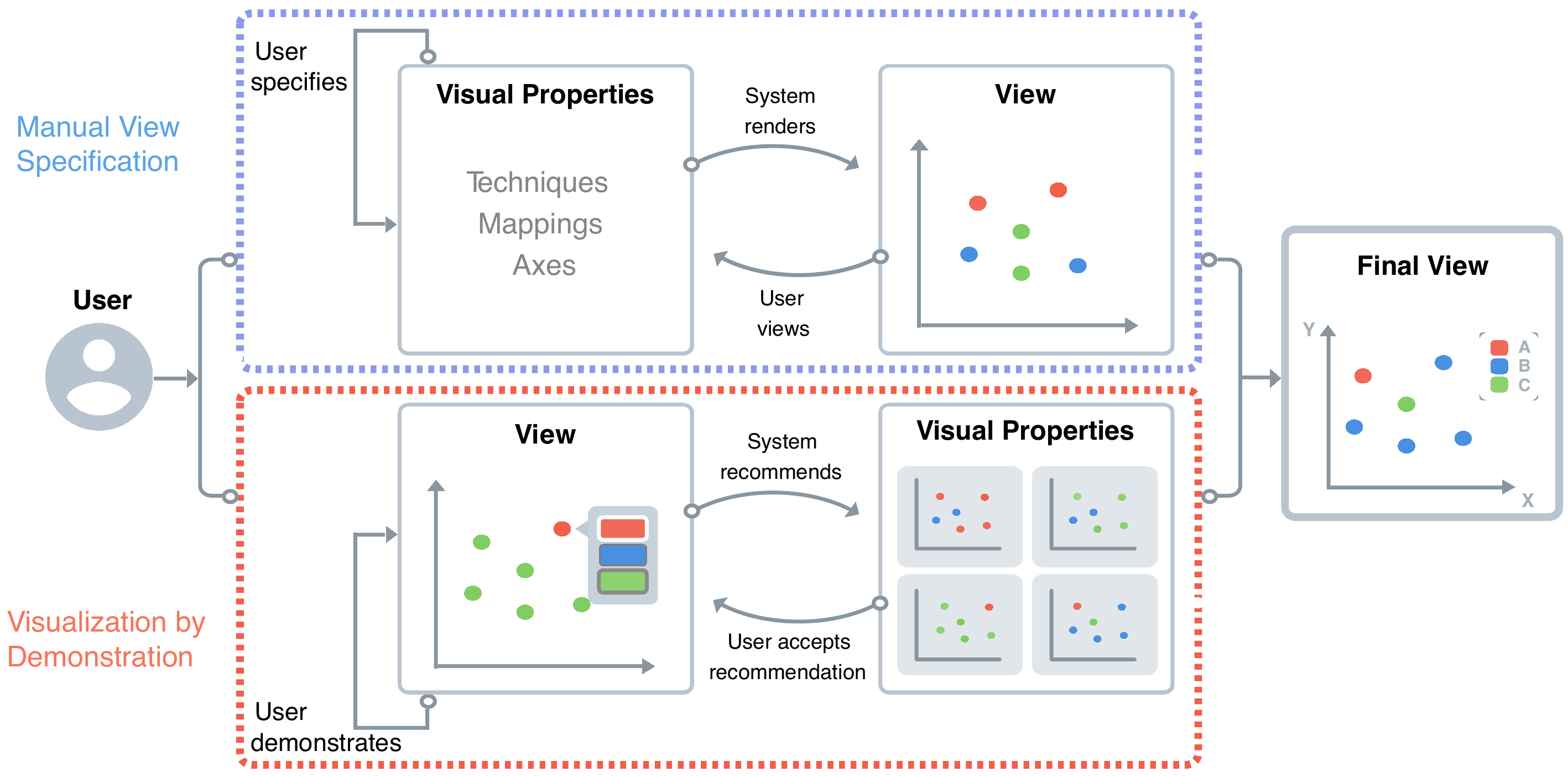}
	\vspace{-1.5\baselineskip}
	\caption{The processes for constructing visualizations using Manual View Specification (MVS) and Visualization by Demonstration (VbD). Figure from~\cite{saket2018evaluation}, used with permission. }
	\label{fig:InteractionModel}
\end{minipage}

\vspace{1.5\baselineskip}

\begin{minipage}[t]{1.0\linewidth}
	\centering
	\includegraphics[width=1.0\linewidth]{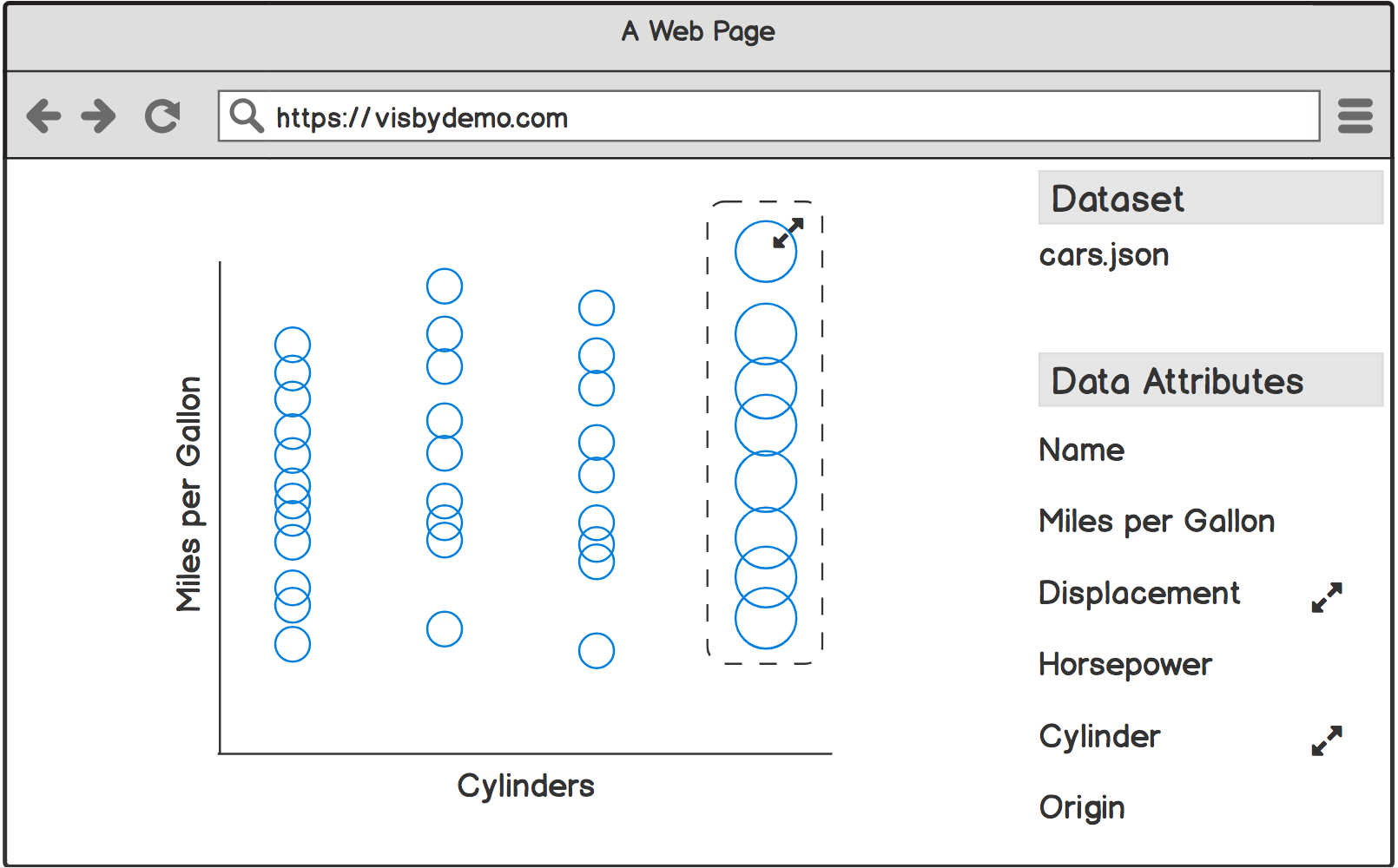}
	\vspace{-1.5\baselineskip}
	\caption{A possible implementation of VbD. 
     Here, a user directly manipulates the size of a set of data points (making them larger), to demonstrate their interest in mapping a data attribute to size. 
     The system then extracts a set of data attributes to map to size and suggest them by showing the resize icon (\faExpand) besides these appropriate data attributes.}
	\label{fig:VBD}
\end{minipage}

\vspace{-1.0\baselineskip}

\end{figure}

In contrast to MVS, VbD~\cite{saket2018evaluation, saketVbD, saket_CGA} does not follow the sequential approach to first mapping data attributes to visual properties, then rendering the views.
Instead, VbD lets people provide partial demonstrations of their intent at the View level through direct manipulation of visual marks~\cite{saket:2017:evaluating, saket:encoding2019}.
Then, the system infers lower-level specification and suggests or applies changes at the view level.
For example, users could resize a few data points to convey their interest in mapping size to a data attribute. 
In response, the system extracts data attributes that can be mapped to size and suggests them (see Figure~\ref{fig:VBD}).

\renewcommand{\textfraction}{0}
\renewcommand{\dblfloatsep}{0pt}

VbD builds on explorations of demonstration-based approaches in related computing areas. 
Notably, programming by demonstration~\cite{cypher1993watch} enables users to generate code by providing visual demonstrations of intended software functionality. Programming by demonstration incrementally improves the state of the system by continuing to demonstrate further changes or by directly editing the produced code. 
Other domains that have successfully used the ``by demonstration'' paradigm include 3D drawing by demonstration~\cite{Igarashi:drawing}, data cleaning by demonstration~\cite{Lin:dataCleaning}, and interactive database querying by demonstration~\cite{Zloof1975QueryBE}.

\begin{figure*}[t]
  \centering
  \includegraphics[width=.96\linewidth]{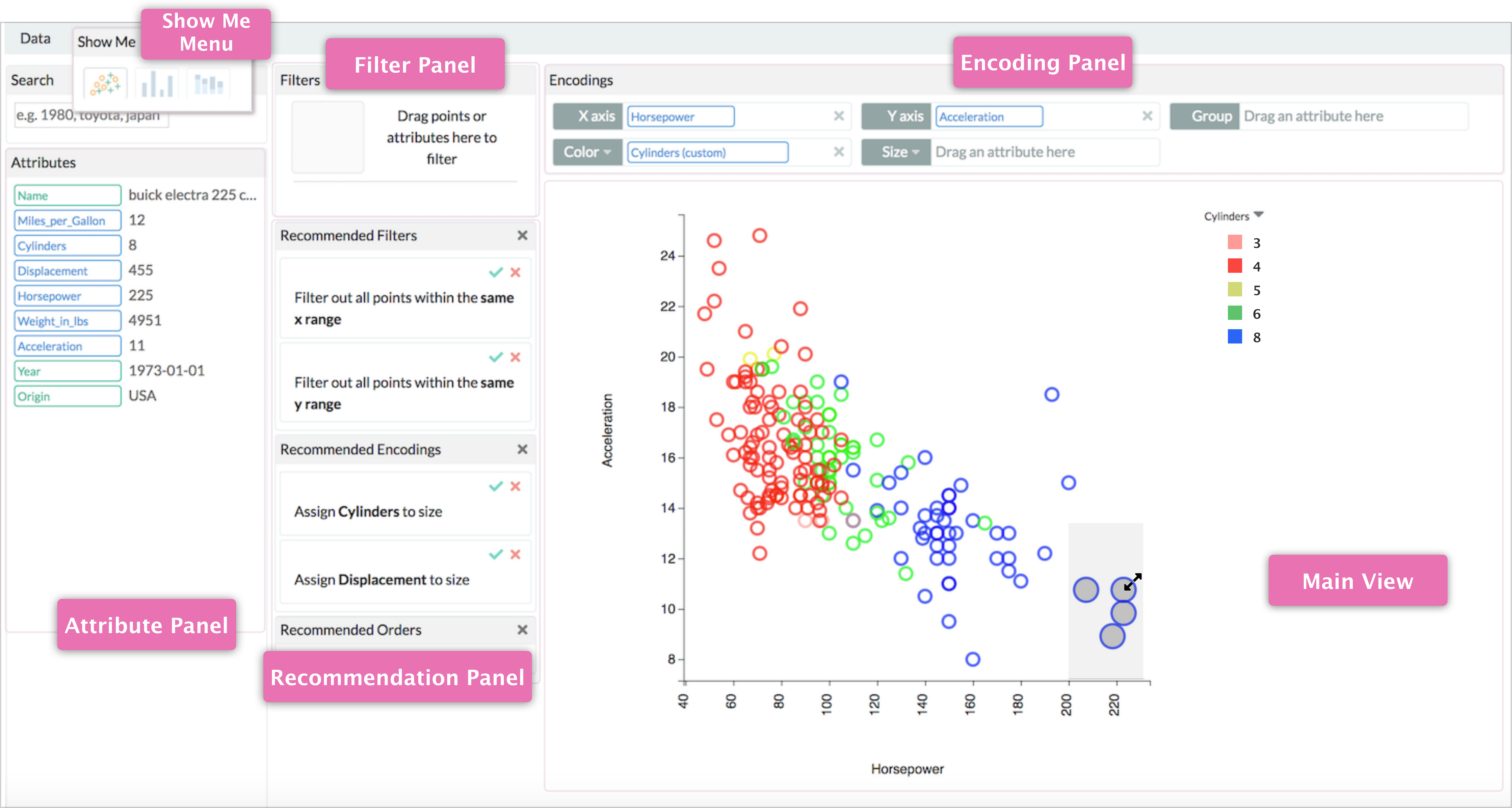}
  \vspace{-0.5em}
  \caption{
  \tool consists of the following elements.
  The \textit{Show Me} Menu shows the types of visualizations the tool supports.
  The \textit{Attribute} Panel shows the data schema, listing all attributes in the dataset. 
  The \textit{Filter} Panel shows user-specified filters. Filters are created through drag and drop of data attributes or data points into the panel.
  The \textit{Recommendation} Panel shows suggestions made by the system in response to a demonstrations made by the user.
  The \textit{Encoding} Panel contains placeholders for different visual encodings. Users can map data attributes to visual encodings by dragging and dropping data attributes onto encoding placeholders.
  The \textit{Main View} contains the visualization. 
  }
	\label{fig:teaser} 
    \vspace{-0.6em}
\end{figure*}

\subsection{Studying the Effectiveness of Interaction Paradigms}

Previous research has built an understanding of advantages and disadvantages of different paradigms supported in visualization tools. 

Grammel et al.~\cite{grammel2010information} studied how visualization novices construct visualizations with a shelf configuration software (MVS). They found that participants in their study had difficulties in breaking down their questions into a set of lower level operations to perform on the interface. 
We hypothesize this is because specifying the visual properties prior to seeing the visualization is a challenging task for users with little visualization expertise. 
Another study revealed differences between top-down or bottom-up visualization tools in terms of how visualization novices construct visualizations and make design choices~\cite{Mendez:2017}. 
Results from this study reveal trade-offs between top-down and bottom-up approaches to creating visualizations. In particular, while top-down approaches result in faster data exploration, bottom-up approaches result in more active data exploration processes thus enable users to better reflect on their data, insights, uncertainties, and open questions.

Previous work also found that mixed-initiative systems that combine both breadth and focus approaches help analysts engage in both open-ended exploration and targeted question answering, because it increases data field coverage compared to a traditional focused approaches~\cite{2017-voyager2}. 
Directly comparing how people construct visualizations with a MVS tool and with a VbD tool~\cite{saket2018evaluation} showed that each interaction paradigm is more efficient than the other for different visualization operations.

In summary, previous research tells us that: 
1) individual interaction paradigms have advantages and disadvantages, 
2) mixed-initiative approaches can increase data analysis coverage,
and 3) interaction paradigms can be complementary (in particular MVS and VbD). 
The next steps in this line of research are: 
i) to understand the feasibility of combining multiple paradigms into a visualization tool that would leverage the benefits of each paradigm -- which we address by designing \tool; 
and ii) to build an understanding of how people might benefit from the availability of multiple paradigms -- which we address through a qualitative study where participants used \tool to explore data. 


\section{Differences Between MVS and VbD}
Although both MVS and VbD offer iterative processes for creating visualizations, they have fundamental differences~\cite{saket2018evaluation}. 
One way to look at these differences is to consider the dimensions of \textbf{visualization construction model} and \textbf{number of intermediary interface elements}~\cite{saket2018evaluation}. 
In terms of the \textbf{visualization construction model},
MVS requires people to specify visualization techniques, 
mappings from data attributes to visual encodings, and other visualization parameters. 
In contrast, VbD requires people to provide visual demonstrations of incremental changes to the visualization. 
It then recommends potential visualization techniques, mappings and parameters based on the system's interpretation of the demonstrations. 
In terms of \textbf{number of intermediary interface elements}, 
MVS introduces interface elements (or instruments~\cite{beaudouin:2000:instrumental}) such as menus and dialog boxes that act as mediators between the user and the visual representation. 
In contrast, VbD lets people interact directly with the visual representation rather than external interface elements, as much as possible. 
Although implementations of VbD usually rely on some external interface elements, for example for accepting or rejecting the recommendations, the number of required interface elements is smaller than with MVS. 

MVS and VbD are also different when considering the dimensions of \textbf{agency} and \textbf{granularity}~\cite{Mendez:2018}.
\textbf{Agency} refers to \textit{who} is responsible for carrying out the visualization construction process: the user or the tool. 
For example, Watson Analytics~\cite{watson-analytics} gives full agency to the tool while iVoLVER~\cite{Mendez:2017} gives full agency to the user. 
With most MVS tools such as Polestar, Many Eyes, and Spotfire, the agency tends to be more on the user side than on the tool side~\cite{Mendez:2018} because design decisions are mostly driven by the user. 
With VbD, agency is shared between the user and the tool because of the automation that occurs as part of recommending visualizations based on user demonstrations. 
\textbf{Granularity} refers to the level at which the tool enables the manipulation of both data and visual representations. 
Most MVS tools have a \textit{coarse} granularity~\cite{Mendez:2018}, as they let users operate on data attributes and manipulate groups of marks (e.g., LARK~\cite{Tobiasz:2009}). 
Tools that have a \textit{fine} granularity like iVolver~\cite{Mendez:2017} let users access individual data values and manipulate individual marks.
VbD has a finer level of granularity than MVS, as it lets users directly manipulate individual graphical encodings rather than attributes.

\section{\tool Walk-through}\label{sec:walkthrough}
We start by illustrating the multi-paradigm functionality of \tool through a scenario. 
Suppose Amy is interested in buying a car that would best match her needs and preferences. She opens \tool and loads the car dataset~\cite{henderson1981building}. The dataset contains 250 cars, each described through 9 attributes such as number of cylinders and miles per gallon.

Amy first wants to get an idea of the relationship between number of cylinders and miles per gallon. 
For that, she uses the MVS paradigm. She drags the \textit{Cylinders} and \textit{Miles per Gallon} attributes from the \textit{Attributes} panel (left-most panel in Figure~\ref{fig:teaser}) and drops them onto the $x$ and $y$ axis placeholders in the \textit{Encodings} Panel (top-right panel in Figure~\ref{fig:teaser}). She finalizes her first visualization by selecting the bar chart under the \textit{Show Me} menu (top-left menu in Figure~\ref{fig:teaser}).

\begin{figure}[t!]
 \centering
   \includegraphics[width=0.93\columnwidth]{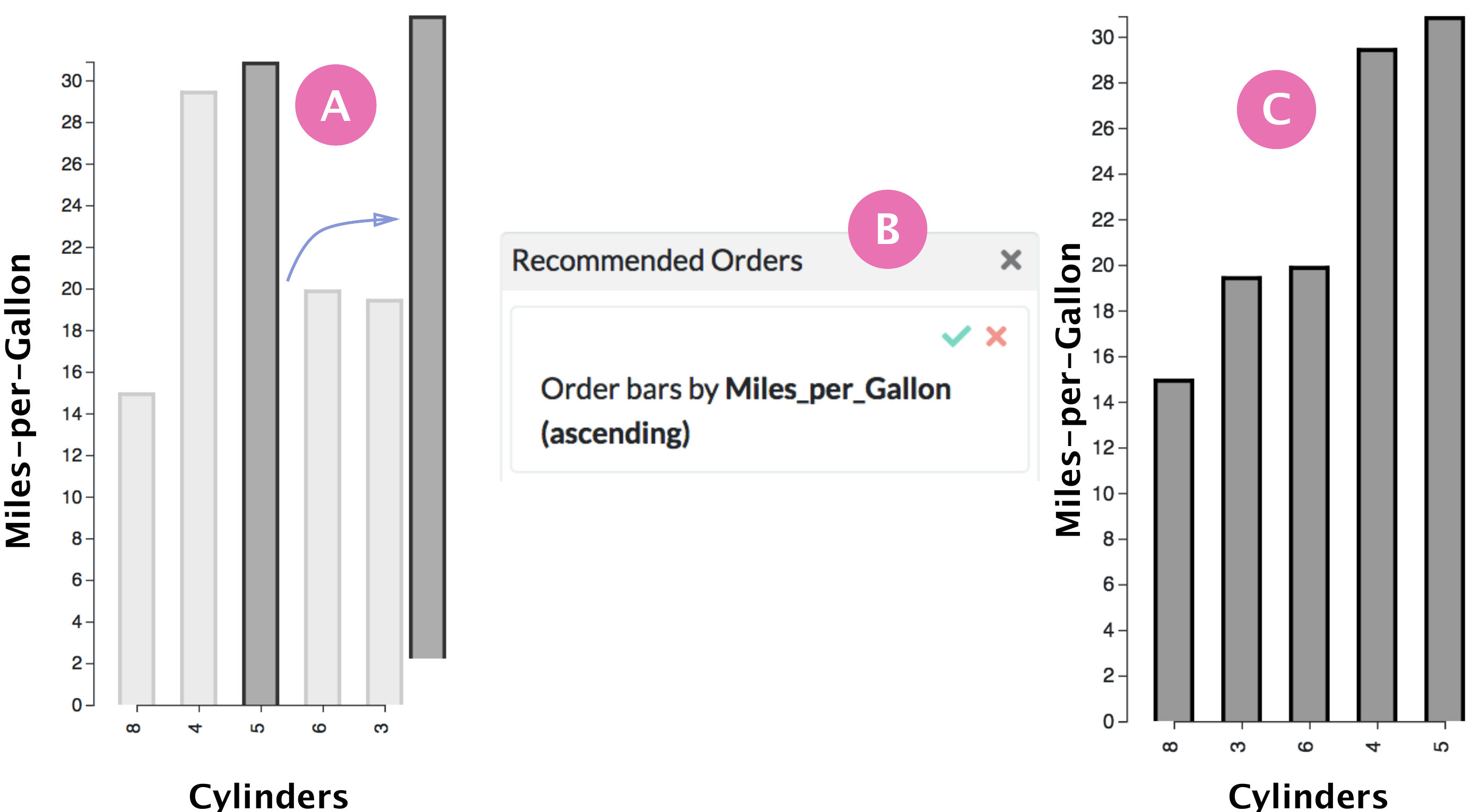}
   \vspace{-0.5em}
 \caption{
 After Amy drags the tallest bar to the extreme right of the bar chart \textbf{(A)}, the system recommends sorting the bar chart either by \textit{Horsepower} or \textit{Miles Per Gallon} in an ascending order \textbf{(B)}. 
 Amy accepts the recommendation to sort by \textit{Horsepower} \textbf{(C)}.
     }
 \label{fig:b}
 \vspace{-0.1em}
 \end{figure}

Amy notices that the average miles per gallon of cars varies according to number of cylinders. 
So she decides to sort the bar chart to see which numbers of cylinders have the highest and lowest average miles per gallon. For that, she uses the VbD paradigm. She selects the tallest bar and drags it the the extreme right of the bar chart to demonstrate her interest in sorting (see Figure~\ref{fig:b}-A). In response, the \textit{Recommendation Panel} is updated based on the system's interpretation of Amy's demonstration (see Figure~\ref{fig:b}-B). Amy accepts the recommendation to sort the bar chart by \textit{Miles per Gallon} in an ascending order; the system updates the bar chart accordingly (see Figure~\ref{fig:b}-C).

\begin{figure}[t!]
 \centering
   \includegraphics[width=0.93\columnwidth]{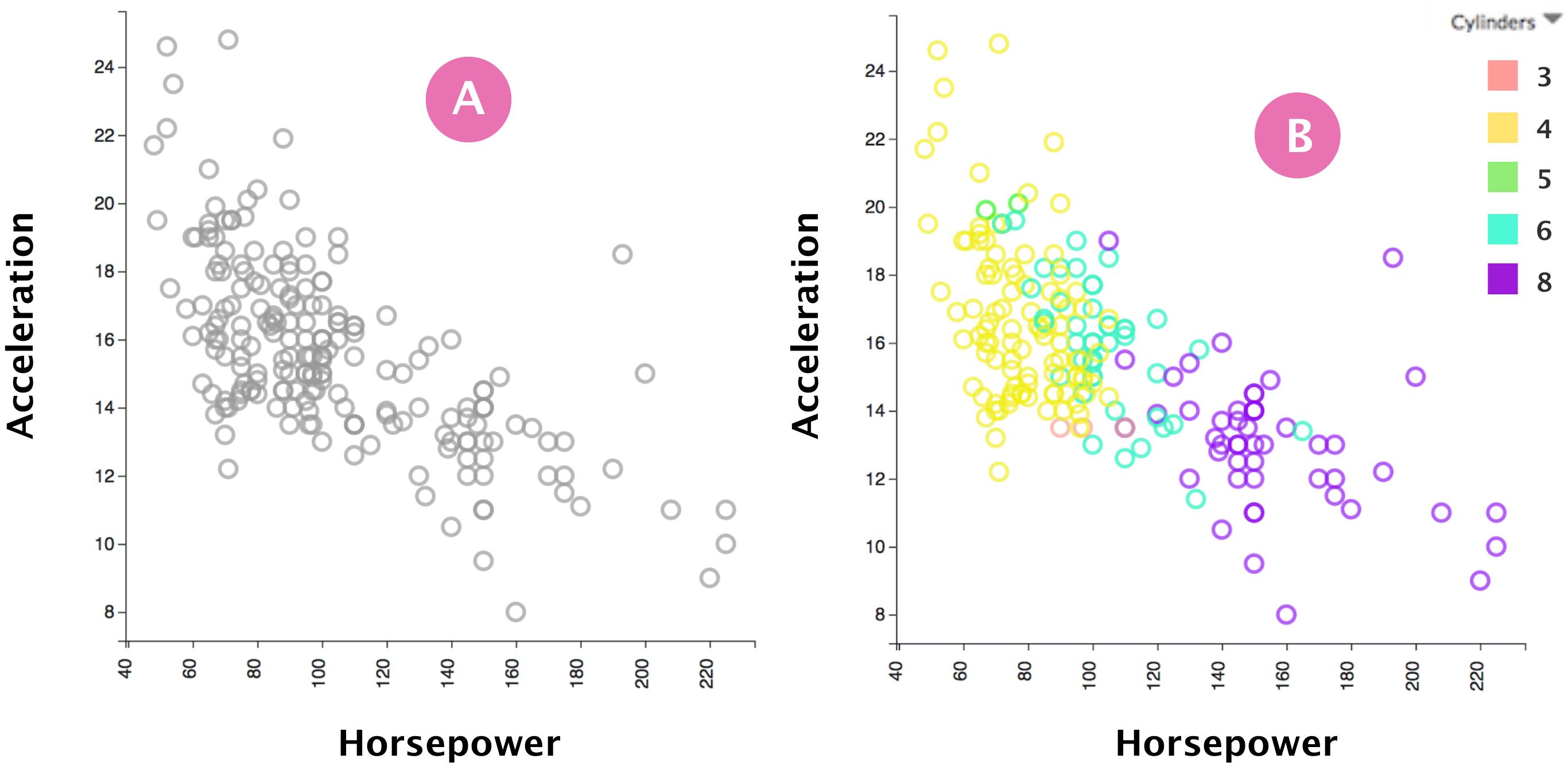}
   \vspace{-0.5em}
 \caption{Amy creates a scatterplot with \textit{Acceleration} on the $x$ axis and \textit{Horsepower} on the $y$ axis \textbf{(A)}, then maps \textit{Cylinder} to color hue \textbf{(B)}.}
 \label{fig:c}
 \vspace{-0.5em}
 \end{figure}

Amy realizes i) that there is not a straightforward relationship between the number of cylinders and the miles per gallon rating; and ii) that the bar chart is not a good visualization for helping her look at individual vehicles.
Thus, she decides to look at the relationships between other dimensions (\textit{Horsepower}, \textit{Acceleration}, and \textit{Cylinders}) using a scatterplot.
She uses the MVS paradigm to switch from the bar chart to a scatterplot. She drags and drops the \textit{Horsepower} attribute onto the $x$ axis placeholder and the \textit{Acceleration} attribute onto the $y$ axis placeholder (see Figure~\ref{fig:c}-A). 
Then she maps the \textit{Cylinders} attribute to color hue using the same drag-and-drop technique (see Figure~\ref{fig:c}-B). 

\begin{figure}[!t]





\begin{minipage}[t]{1.0\linewidth}
	\centering
	\includegraphics[width=0.93\linewidth]{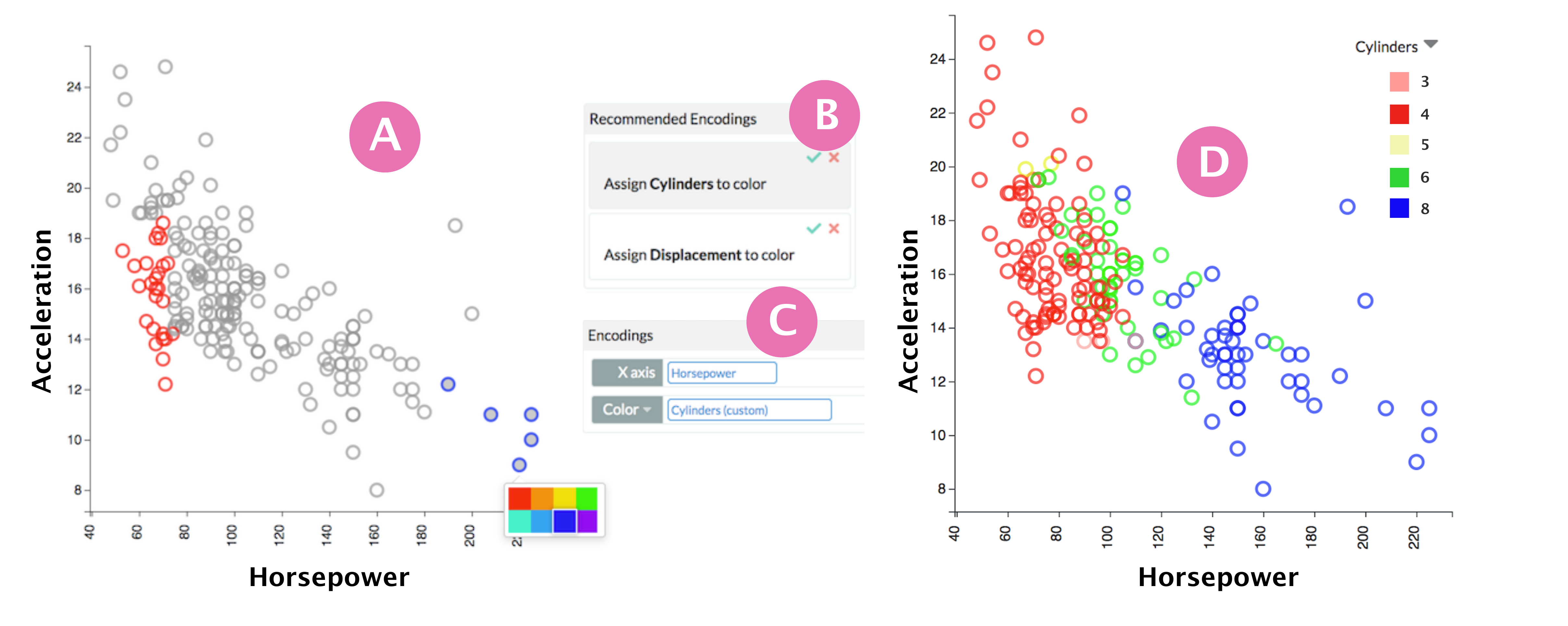}
	\vspace{-0.5\baselineskip}
	\caption{
	    Amy colors a few 4-cylinder cars red and a few 8-cylinder cars blue \textbf{(A)}. 
        The system recommends to map \textit{Cylinder} or \textit{Displacement} to color \textbf{(B)}. 
        Amy accepts the \textit{Cylinders} mapping. 
        This updates the \textit{Encoding} panel \textbf{(C)} and the \textit{Main View} \textbf{(D)}, using the colors she specified.
	}
	\label{fig:d}
\end{minipage}

\vspace{1.5\baselineskip}

\begin{minipage}[t]{1.0\linewidth}
	\centering
	\includegraphics[width=0.93\linewidth]{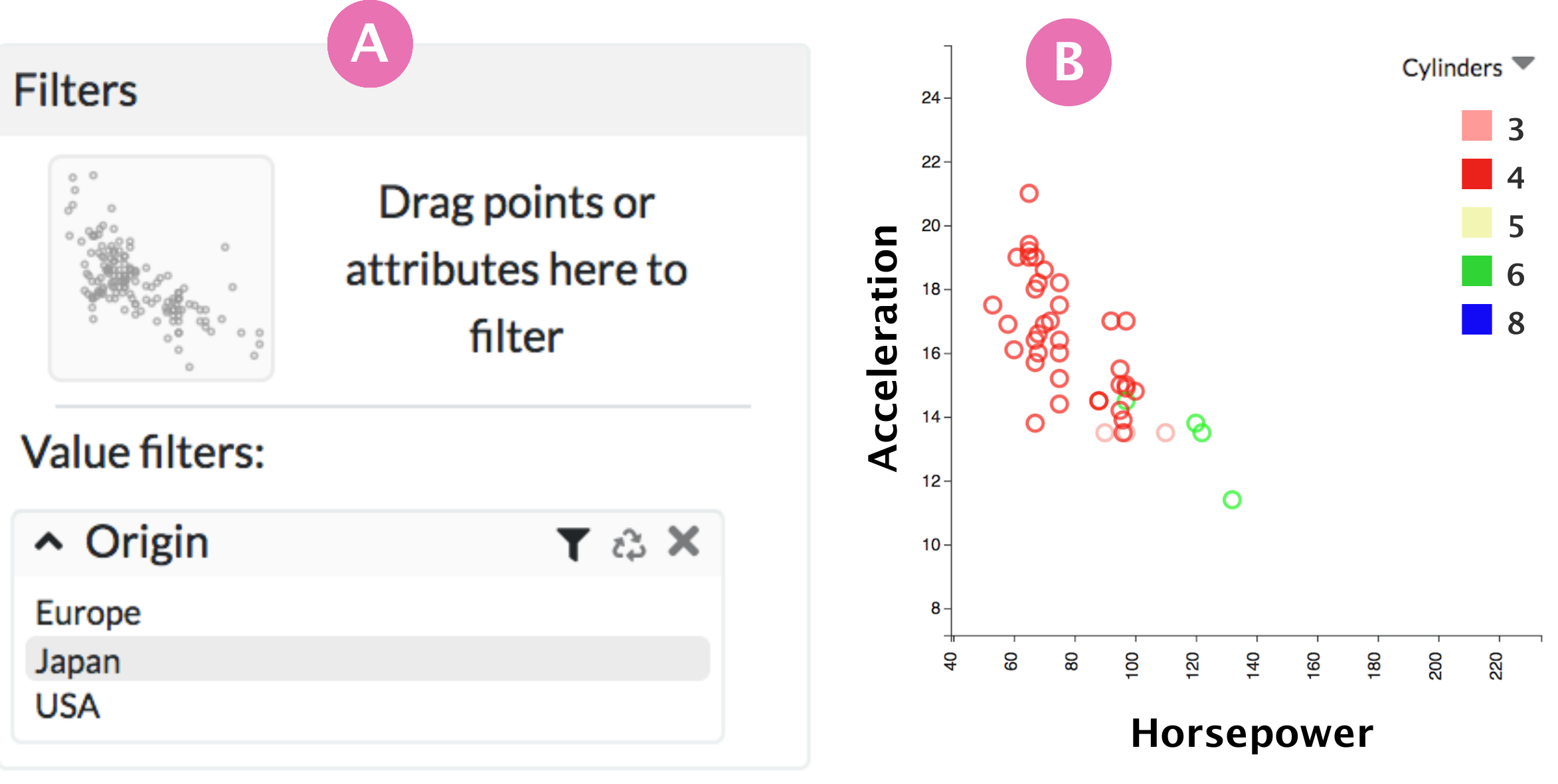}
	\vspace{-0.5\baselineskip}
	\caption{
	    Amy creates a filter to filter out European and American cars \textbf{(A)}. The \textit{Main View} updates to only show Japanese cars \textbf{(B)}.
	}
	\label{fig:e}
\end{minipage}

\vspace{1.5\baselineskip}

\begin{minipage}[t]{1.0\linewidth}
	\centering
	\includegraphics[width=0.93\linewidth]{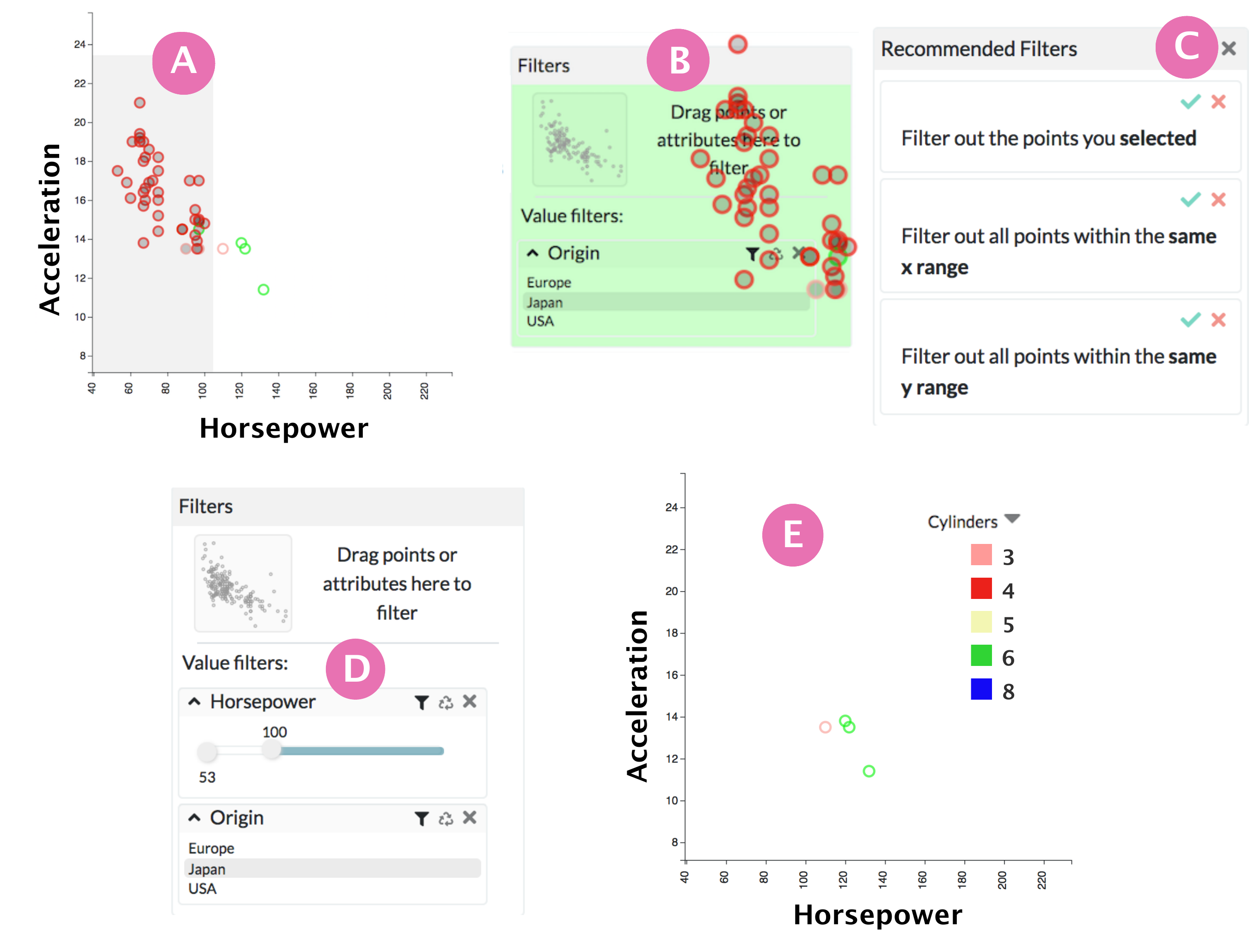}
	\vspace{-0.5\baselineskip}
	\caption{
	    Amy draws a rubber-band rectangle to select cars with low \textit{Horsepower} \textbf{(A)}.
        She drags the selected cars and drops them onto the \textit{Filter} panel to demonstrate her interest in filtering out these cars \textbf{(B)}. 
        Amy chooses to filter out the cars with \textit{Horsepower} within the selected range.
        She then uses the range slider to filter out cars with \textit{Horsepower} below 100 \textbf{(D)}.
        Amy ends up with four similar cars to choose from \textbf{(E)}. 
	}
	\label{fig:f}
\end{minipage}

\vspace{-1.0\baselineskip}

\end{figure}

Amy is not very fond of the color scheme automatically assigned to \textit{Cylinders}. She removes the color mapping, reverting the \textit{Main View} to the one shown in Figure~\ref{fig:c}-A. She decides to use the VbD paradigm to create a color scheme that is more her taste. To demonstrate her intent to customize the color palette, she selects and re-colors a few 4-cylinder cars red and a few 8-cylinder cars blue (see Figure~\ref{fig:d}-A). The system automatically extracts data attributes that can be mapped to color (in this case \textit{Cylinders} and \textit{Displacement}) and recommends them (see Figure~\ref{fig:d}-B). 
Amy accepts mapping \textit{Cylinders} to color. 
The \textit{Encoding} panel now shows the \textit{Cylinders (customized)} attribute on the color placeholder (see Figure~\ref{fig:d}-C) and the color of the data points is updated according to the new color scheme (see Figure~\ref{fig:d}-D).



Amy remembers a friend of her's mentioned that Japanese cars have low fuel consumption. She decided to explore this notion, and uses the MVS paradigm to exclude non-Japanese cars, by dragging and dropping the \textit{Origin} attribute onto the \textit{Filter} Panel (see Figure~\ref{fig:e}-A). 
The \textit{Filter} panel now shows the three values for the \textit{Origin} attribute. 
She excludes the European and American cars by deselecting these in the filter she created (see Figure~\ref{fig:e}-A). 
This updates the overview of the filtered points in the \textit{Filter} panel and the \textit{Main View} (Figure~\ref{fig:e}-B).


While low consumption is important, Amy does still want some sporty aspects to her driving experience, and thus would like a car with high \textit{Horsepower}.
She uses the VbD paradigm to lasso-select cars with \textit{Horsepower} below 100 (see Figure~\ref{fig:f}-A).
She demonstrates her interest in filtering out the selected cars by dragging the them out of the \textit{Main View} and dropping them onto the \textit{Filter} panel (see Figure~\ref{fig:f}-B). 
Amy explores the different recommended options (see Figure~\ref{fig:f}-C) by hovering over them on the Recommendation Panel, which provides a preview of the change in the main view.
She accepts the recommendation to filter out all points within the same $x$ range. 
As a result of this, the system automatically creates a new filter for the \textit{Horsepower} attribute in the \textit{Filter} panel. Because the data is quantitative, the filter provides a range slider for Amy to fine-tune the filtering criteria (see Figure~\ref{fig:f}-D). 
The \textit{Main View} updates to reflect the new filter (see Figure~\ref{fig:f}-E). 
Amy notices that one of the four cars (in red) has less cylinders than the other three (in green). Given that the cars are similar otherwise, she thinks that the \textbf{Mazda MX-5} is her best option. After narrowing down her options to only 4 cars, she decides to go test drive each of them.


\section{Preliminary Study}
Designing a tool that combines paradigms (MVS and VbD in our case) is a challenging endeavor. 
We faced several design decisions, including: 
\textit{Should each operation be supported by only one paradigm (perhaps the one best suited for the operation) or by both paradigms? Should both paradigms work in conjunction or independently?} 
Thus, we started with a preliminary study to better understand the design space of a multi-paradigm visualization tool by observing the usage patterns and the difficulties people may have for each paradigm.

We first designed a prototype that supports a variety of operations for scatterplots and bar charts. 
These operations include mapping data attributes to axes and mark properties (e.g., size and color), 
switching from one visualization technique to another, 
filtering data points, and sorting according to an axis. 
Both MVS and VbD had full coverage of the available operations. 
For instance, to sort a bar chart, one could either: 
i) with MVS, click the sort button on the control panel (similar to Tableau); 
or ii) with VbD, drag the shortest/tallest bar to extreme left or right to demonstrate their intent in sorting the bar chart. 
The system would then interpret the intent and suggest sorting the bar chart.

We recruited four participants (3 male, 1 female). 
We first explained the tool and available interactions. 
Then, we asked participants to imagine their employer asked them to analyze a dataset about movies (the Movies dataset~\cite{TableauData}) 
using the tool for 20 minutes, and to report their findings about the data. 
We encouraged participants to try both paradigms and to verbalize their thought process while exploring the data. 
Results from this preliminary study emphasized three patterns for constructing visualizations using a multi-paradigm tool:

\ourParskip
\noindent \textbf{Pattern 1: Try it out first.} Participants performed the same operation using both paradigms one after the other when they first started interacting with the system. Trying out the different paradigms helped them better understand the system and possible interactions.

\ourParskip
\noindent \textbf{Pattern 2: Assess and choose.} Over time, participants' choice of paradigm converged toward using the paradigm they found to be most efficient for a given operation. 
For instance, all four participants preferred using MVS to switch between visualizations as they found this easier and more efficient than using VbD. 
For some operations like filtering data points or mapping a data attribute to color, they found the two paradigms to be equally effective and used both interchangeably.

\ourParskip
\noindent \textbf{Pattern 3: Combine.} Participants sometimes combined MVS and VbD to complete a single operation. 
For example, a participant first used MVS to map a data attribute to color in a scatterplot, by dragging and dropping the data attribute to the color shelf. 
Then, she said: \textit{``I don't like the colors''}, and used VbD to manually color a few data points, representative of a custom color palette she had in mind. The system then recommended color mappings based on these colors. 

\ourParskip
Based on these three patterns, we iterated over our initial design to develop a new version of our tool, called \tool.

\section{\tool}
We designed and developed \tool (see Figure~\ref{fig:teaser}) as a proof of concept multi-paradigm prototype that blends two interaction paradigms (MVS and VbD).
We implemented \tool using JavaScript, TypeScript, and D3~\cite{d3:infovis11}. 
It is available at {\small\url{https://github.com/liger-vis/LigerPrototype}}.

\subsection{Design Goals}\label{sec:desginGoals}

Based on the findings from our preliminary study and lessons learned from previous work~\cite{grammel2013survey, Mendez:2018, beaudouin:2000:instrumental, saket2018evaluation,Mendez:2017}, we introduce three general design goals for a multi-paradigm tool.


\begin{enumerate}[label=$\bm{G_\arabic*$},align=left,itemsep=0.33em,labelsep=4pt,leftmargin=*]

\item\label{dg:1} \textbf{Not every operation must be supported by both paradigms.}
Previous work indicates that the effectiveness of interaction paradigms varies depending on the operation at hand~\cite{saket2018evaluation,Mendez:2017}. 
Results from our preliminary study (\textbf{Pattern 2: Assess and choose}) indicate that
people choose an interaction paradigm for a given operation mainly based on its effectiveness for that operation. 
Applying this design goal should lead to a simpler user interface, fewer conflicts between paradigms, and an enhanced user experience. 

\item\label{dg:2} \textbf{Paradigms should work in conjunction rather than independently.} 
Results from our preliminary study (\textbf{Pattern 1: Try out first} and \textbf{Pattern 3: Combine}) indicate that for operations supported using two paradigms, users find it desirable to have the freedom to switch between paradigms, including times where doing so completes a single operation. 
Applying this design goal should result in multiple interaction paradigms to work hand in hand, and allow users to seamlessly switch between paradigms \textbf{anytime} during their visualization construction process.

\item\label{dg:3} \textbf{Facilitate synchronizing changes between paradigms.}
Results from our preliminary study (\textbf{Pattern 3: Combine}) indicate that when users perform an operation using one paradigm, the system should show the corollary interactions with the other paradigms. For example, if one filters a set of points using VbD, the system can show a range slider highlighting the range of values that are filtered, that can be further used using MVS. 
This design goal helps users understand how multiple paradigms support the same operation, and indicates how an operation can be supplemented with a second paradigm (e.g., fine-tuning a filter operation using MVS). 

\end{enumerate}

\subsection{The \tool Interface}
We introduced \tool's interface through the walk-through in Section~\ref{sec:walkthrough}, and Figure~\ref{fig:teaser} provides a description of each element in the interface.
Here we describe how \tool supports MVS and VbD, and how it implements the design goals specific to a multi-paradigm tool.

\inlinebold{How \tool supports MVS}
There are two main guidelines for designing MVS tools~\cite{grammel2010information,ahlberg1996spotfire}.
\textit{First}, the tool must support mapping data attributes to various visual encodings.
\tool implements this guideline through a shelf-configuration design (i.e., dragging the data attributes and dropping them onto encoding shelves in the interface) -- like e.g., Tableau~\cite{Tableau} and Polestar~\cite{PoleStar}. 
\textit{Second}, the tool must update the visualization in real time after visual encodings are created or updated.
\tool renders the Main View every time a visualization property is specified.

\inlinebold{How \protect\emph{\tool} supports VbD}
There are three main guidelines for designing VbD tools~\cite{saketVbD}.
\textit{First}, the tool must enable direct manipulation of visual representation as a method for providing visual demonstrations. 
In \tool, the Main View renders the visualization and allows users to provide visual demonstrations by manipulating graphical encodings of the visualization itself. 
\textit{Second}, the tool must balance the human and machine workload in the visualization construction process. 
\tool suggests possible relevant visual transformations in response to given demonstrations. 
\textit{Third}, the tool must enhance the interpretability of recommendations.  
\tool implements this guideline in several ways: 
i)  the Recommendation Panel organizes recommendations in different divisions based on their types (e.g., Recommended Filters, Recommended Encodings);
ii) hovering over a recommendation provides a preview (feedforward~\cite{norman2013design,VermeulenFeedforward}) of how the visualization would be updated;
and iii) each recommendation is explained in natural language. 

\inlinebold{How \tool supports multiple paradigms}
Each interaction paradigm implemented in \tool supports a subset of the available operations (\ref{dg:1}). 
For example, both previous research~\cite{saket2018evaluation} and our preliminary study tell us that switching between visualization types is non-trivial using VbD. 
Therefore, VbD does not support this operation in \tool.
Both MVS and VbD support the operations for which they are well-suited.

In \tool, MVS and VbD work in conjunction, meaning that users can switch between paradigms anytime during the visualization construction process (\ref{dg:2}). 
For example, one can first create a scatterplot and assign a data attribute to the color of the points using MVS. 
They can then continue their construction process by mapping a data attribute to the size of the points using VbD. 
\tool also enable users to switch between paradigms to complete a single operation. 
For instance, one can first use VbD to filter out a specific set of points, then use a range slider to fine-tune the filtering criteria using MVS.

When performing an operation using a particular paradigm, 
\tool synchronizes changes between paradigms by showing the corollary interaction with another paradigm (\ref{dg:3}). 
For example, if a user maps a data attribute to color using VbD, 
the system automatically updates the color placeholder on the Encoding Panel to show the equivalent of that interaction using MVS. 
Alternatively, if a user filters out a subset of data points using VbD, 
the system automatically shows the filtering criteria using graphical widgets 
(range sliders or check boxes depending on data attribute type) 
on the Filtering Panel (see Figure~\ref{fig:f}).


\subsection{The \tool Architecture}
To blend MVS and VbD, \tool employs two main components: the \textbf{Interface} and the \textbf{Recommendation Engine}. 
Below we describe their specific functions and how they communicate with each other. 

For MVS, interactions (e.g., dragging data attributes and dropping them to the shelves or selecting the visualization type from the ``Show Me'' menu) are handled by the Interface. 
Once users specify visualization properties such as color of data points, 
the system receives the users' specifications and updates the visualization shown on the Main View. 
\tool supports different visualization construction operations using MVS, like existing tools such as Tableau and Polestar. 

For VbD, \tool's architecture follows the same pipeline as VisExemplar~\cite{saketVbD}. 
Users' visual demonstrations (e.g., selecting, dragging, coloring, and resizing data points) are handled by the Interface. 
Each demonstration is a set of actions that a user takes to show parts of the expected results visually. 
When a user provides a demonstration, the interface calls a set of intent functions in the Recommendation Engine. 
Intent functions are a set of rules that predict the potential meaning(s) of the given demonstration (i.e., they guess user intent). 
The recommendation engine then computes and ranks the changes that can be applied to the visualization given the predicted meaning(s). 
The Recommendation Engine sends these potential changes to the Interface, 
that is responsible for showing the recommendations to the user, 
including what to show, when to show it, and how to show it.

\subsection{Operations Supported in \tool}
Table~\ref{TAB:operations} provides the five operations supported by \tool according to visualization type and interaction paradigm.
In this section, we explain how MVS and VbD support each operation.

\inlinebold{\opAxis: Map Data Attributes to Axes}
Only MVS supports this operation. 
Users drag a data attribute from the Attribute Panel and drop it onto one of the shelves on the Encoding Panel (see Figure~\ref{fig:teaser}).

\inlinebold{\opMark: Map Data Attributes to Mark Properties}
With MVS, users map a data attribute to color or size by dragging that data attribute and dropping it onto the color or size encoding placeholders. 
With VbD, users first directly manipulate the mark properties for a few data points. 
Then, \tool will recommend data attributes that can be assigned to those properties. 
For example, in a bar chart, users could color the bars based on their values. In response, the system would recommend potential data attributes that can be mapped to the color encoding.

\inlinebold{\opSwitch: Switch Between Visualization Types}
Only MVS supports this operation, via the ``Show Me'' menu (top-left menu in Figure~\ref{fig:teaser}).

\inlinebold{\opFilter: Filter Out Data Points}
With MVS, users drag a data attribute and drop it onto the Filter Panel. 
The system then shows the range of filtered values on the Filter Panel. 
\tool shows the filtered values differently depending on the data attribute type (e.g., quantitative, categorical). 
For instance, for a quantitative data attribute, the system will show the filtered values through a range slider that can be further tuned. 
With VbD, users demonstrate their interest in filtering data points by selecting some of those points then dragging and dropping them from the Main View onto the Filter Panel. 
In response, \tool suggests ways to specify the selection of points to filter, as illustrated in Figure~\ref{fig:f}.

\inlinebold{\opSort: Sort the order of Bars}
With MVS, users change the order of the bars by clicking on the sort bars buttons. These buttons appear on the top menu when the visualization is either a bar chart or a stacked bar chart. 
With VbD, users demonstrate their interest in sorting a bar chart or a stacked bar chart by dragging the tallest/shortest bar to the extreme left or right side of the visualization. 
In response, \tool recommends sorting the bars in an ascending or descending order.

\begin{table}
\setlength\tabcolsep{3pt}
\renewcommand{\arraystretch}{0.0}
\footnotesize
  \caption{List of operations that \tool supports, according to visualization type and interaction paradigm.} \label{TAB:operations} 
   \begin{tabular}{p{0.28\textwidth}p{0.115\textwidth}p{0.025\textwidth}p{0.025\textwidth}}
    \toprule
     {\sc Operations} & {\sc Visualizations} & {\sc MVS} & {\sc VbD}\\
    \midrule

       \opAxis: Map data attributes to axes 
      &
       \begin{minipage}{.1\textwidth}
          \includegraphics[width=\linewidth]{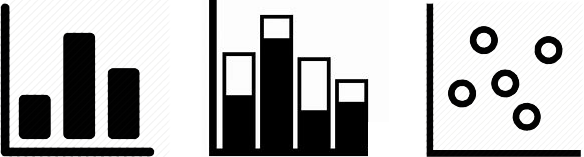}
       \end{minipage}
      &
      \cmark
      &
 \\
         \midrule 
      \opMark: Map data attributes to mark properties 
      &
       \begin{minipage}{.1\textwidth}
          \includegraphics[width=\linewidth]{pictures/threeVis}
       \end{minipage}
      &
       \cmark
      &
       \cmark
 \\
         \midrule 
     \opSwitch:  Switch between visualizations 
      &
       \begin{minipage}{.1\textwidth}
          \includegraphics[width=\linewidth]{pictures/threeVis}
       \end{minipage}
      &
      \cmark
      &
       \\
         \midrule 
      \opFilter:  Filter out data points
      &
       \begin{minipage}{.1\textwidth}
          \includegraphics[width=\linewidth]{pictures/threeVis}
       \end{minipage}
      &
      \cmark
      &
      \cmark
    \\
            \midrule 
    \opSort: Sort the order of bars
      &
       \begin{minipage}{.075\textwidth}
          \includegraphics[width=\linewidth]{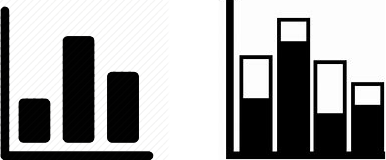}
       \end{minipage}
      &
      \cmark
      &
      \cmark
 \\

    \bottomrule

 \end{tabular}\vspace{-1em}
\end{table}


\section{Evaluating \tool}
We initially considered performing a study to compare the effectiveness of a multi-paradigm tool with a single-paradigm tool (e.g., Polestar or Tableau).
However, our goal is not to examine the benefits of single-paradigm versus multi-paradigm; 
it is to understand \textit{how} participants use each of the paradigms blended in a unified system, and to study the features and design of \tool.
To address this goal, we conducted a think-aloud exploratory observational study,
i) to understand participants' processes when \textbf{using} \tool (e.g., how often do participants use each interaction paradigm? What types of visualization specifications do participants create using each interaction paradigm?); 
and ii) to reveal \textbf{barriers} of \tool and each paradigm (e.g., when and how difficulties happen). 
The two datasets used in our study, the operations used for training sessions, the study protocol, and the data we collected are available at {\small\url{https://github.com/liger-vis/materials}}.

\subsection{Data Collection Methods}
To answer our research questions, we collected a range of data that capture participants' processes and preferences.
At the beginning of the study, we used questionnaires to collect participant demographic and background information. 
During the main study, we took written notes of participants' interactions with \tool. 
We screen- and audio-recorded the whole study.
We then conducted a semi-structured interview where we asked participants a set of questions to collect their preferences and subjective opinions about the tool and the two paradigms.

\subsection{Participants and Settings}
We recruited 10 non-color blind participants (2 females, 8 males), aged 22--34 (mean 27.8) via email and word of mouth at our university.
None of them had participated in the preliminary study.
They were all undergraduate and graduate students enrolled in computer science (7), Social Science (1) and Mechanical Engineering (2), 
were familiar with reading visualizations, and had created visualizations before. 
Some participants had used tools such as Microsoft Excel (6), D3.js (3), SPSS (3), Tableau Software (2) and Google Charts (1), and with programming languages such as R (5), Python (4) and Matlab (2). 
They sat roughly 30--40 cm from a 13'' LCD display with a resolution of 2560$\times$1600 pixels equipped with a mouse. \tool was shown in full screen. 

\subsection{Datasets}
We used two datasets in our study: 
the Cars dataset for the introduction and training sessions (250 cars, 9 attributes);
and the Movies dataset for the main experiment (335 movies, 12 attributes). 
We selected these datasets because: 
i) participants were likely to be familiar with the meaning of the attributes (e.g., meaning of IMDb rating, profit, genre); 
and ii) the datasets are complex enough in terms of number of data cases and attributes to support an open-ended data exploration task.

\begin{figure*}[ht]
\centering
   \includegraphics[width=\linewidth]{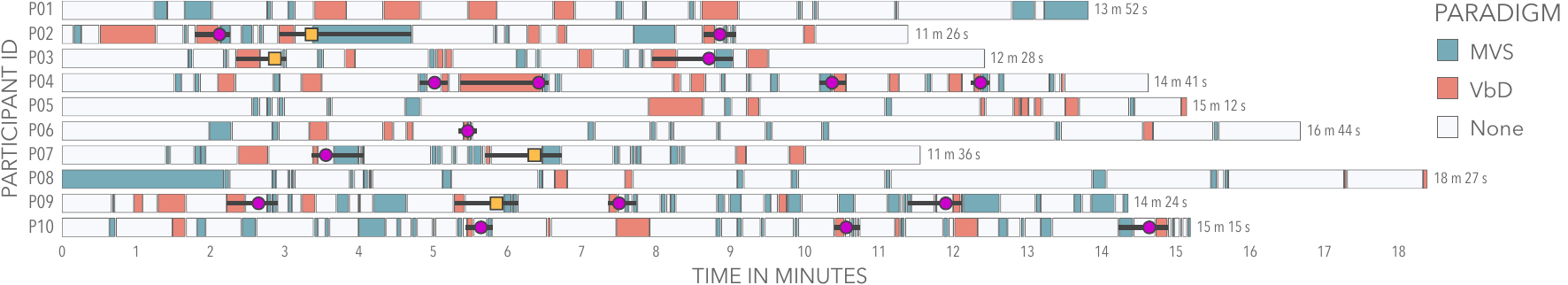}
   \vspace{-1.5em}
\caption{
    The operations participants performed during the \textit{main study} (data exploration). 
    Color indicates which paradigm that was used for each operation. 
    The white spaces indicate when participants were not using any of the paradigms, for instance, when they were hovering over data points or reporting findings about the data. 
    Two symbols indicate when participants switched from one paradigm to the other to perform a single operation, along with a horizontal line that shows the time interval for that operation.
    A \protect\glyph{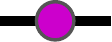} indicates that participants combined the two paradigms to perform a single operation; and
    a \protect\glyph{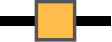} that they switched paradigm because they found that their current paradigm was not effective for that operation.
} 
\label{fig:interactionpatterns} 
\vspace{-1.0em}
\end{figure*}

\subsection{Tasks}
We designed ten \textit{training tasks} that involve the visualization construction operations supported in \tool.
We interacted with the tool ourselves to list all the ways in which it supports visualization construction. 
We obtained a list of 22 operations (e.g., assign a data attribute to the size or color of data points). 
We also reviewed taxonomies of tasks commonly used for interactive visualization construction (e.g.,~\cite{Shneiderman:taxonomy, yi2007toward, Ren:taxonomy, dix1988starting}). We discussed our 22 operations through the lens of these taxonomies to assign each into one of the five categories of operations listed in Table~\ref{TAB:operations}. 
Last, we selected two tasks per category (10 tasks in total), favoring diversity of interactions they involve and coverage of both paradigms. The 10 tasks are detailed in the supplemental material.

The \textit{main task} is a data exploration task where participants are given and goal then try to achieve this goal using \tool.
We opted for an exploratory task because we are interested in the qualitative understanding of how people use a multi-paradigm tool, 
rather than in measuring the performance of people to complete low-level tasks accurately and/or quickly.

\subsection{Procedure}

\vspace{-0.4em}
\inlinebold{1. Introduction (\texttildelow 10 min)}
We briefed participants about the purpose of the study and their rights. 
Then we asked them to fill out the study consent form and the questionnaire on demographics and visualization expertise. 
Next, we gave participants a brief introduction to \tool's UI where we walked them through different features and supported interactions. 
We encouraged them to ask questions during this phase. 

\inlinebold{2. Training (\texttildelow 20 min)}
We gave participants a printed list of the 10 training tasks (2 tasks for each of the 5 categories of operations) to perform on the Cars dataset, in randomized order for each participant. 
We informed participants that we would not measure how quickly they perform tasks, so they should feel free and interact naturally with the tool and ask as many questions as they want. 
However, we told participants that they must complete each training task correctly before moving to the next, that this phase is limited to 15 minutes. 
Once they had completed the 10 tasks, participants could freely interact with the tool for an additional 5 minutes. Then they took a short break.

\inlinebold{3. Main study (\texttildelow 20 min)}
We asked participants to explore the Movies dataset and look for interesting facts about the data. 
Specifically, we told the participants:
\textit{``Imagine you are planning to watch a movie. Given this dataset about Movies, please make a data-driven decision using our tool for 10--15 minutes and come up with a list of movies that you should be watching.''} 
We asked participants to verbalize analytical questions they have about the data, interactions they perform to answer those questions, and their answers to those questions, in a think-aloud manner. 
We also asked them to focus on data-driven findings rather than preconceived assumptions and previous knowledge about the data. 
The participants were could not ask questions during this phase. 
We did not interrupt the participants except to remind them to think aloud.

\inlinebold{4. Follow-up Interview (\texttildelow 10 min)} 
We asked the following questions:

\begin{enumerate}[label=(\alph*)]
\item What do you think are the major obstacles/roadblocks while using the tool to solve your problems? How did you resolve the issue?

\item Tell me about your general experience with this tool.

\item How did you use each of the interaction techniques while exploring your data with this tool? 

\item What were the situations that you found one interaction technique more effective/useful than another one?

\item What do you think are the major obstacles/roadblocks of each of the interaction techniques? How did you resolve the issue?

\item Do you find it sometimes useful to use both interaction techniques together to achieve something? Do you have specific examples when you did it? Or other ideas when it could be useful?
\end{enumerate}

\inlinebold{5. Wrap-up (\texttildelow 5 min)}
The experimenter thanked the participants, who received a \textdollar 10 gift card and were invited to ask additional questions about the study.

\subsection{Data Analysis}
We analyzed the experimenter's notes of participants' interactions and the 251 minutes of screen-capture videos in three phases. 

In the first phase, a researcher watched 5 random videos out of the 10, to obtain a general sense of the data. 
Then the researcher coded all operations for all 10 videos in terms of paradigm used, visualization type, start time and end time of the operation (close coding). 

In the second phase, the researcher went through the videos again to identify the most common and unexpected patterns (open coding). 
During this phase, they mainly concentrated on processes of the participants in terms of \textbf{usage} 
and \textbf{barriers}.
For example, they looked for cases that participants combined both interaction paradigms to complete a single task or cases where participants switched to a different interaction paradigm because of an inefficiency of one paradigm. 

In the third phase, a researcher transcribed the interviews. 
Then they identified the meaningful text segments and assigned a code word or phrase to describe the meaning of the text segment (open coding). 
The coding process was iterative with two passes by a single coder in which the coder developed and refined the codes. 
For example, the codes included phrases such as \textit{``major roadblocks''}, \textit{``strengths''}, and \textit{``combined both paradigms''}. 
Finally, as a team we identified frequently occurring codes to form higher-level descriptions of the results.

\subsection{Study Results}
We first describe when and how participants used each interaction paradigm. We then explain situations where participants preferred using one paradigm over another.

 
\subsubsection{Do people use a multi-paradigm visualization tool?}
As shown in Figure~\ref{fig:interactionpatterns}, \textbf{all participants used both interaction paradigms} during the main phase of the study.
Based on the interview data, participants found it empowering and effective to be able to leverage both paradigms. 
For instance, P9 explained how they used both paradigms to filter points differently: 
\textit{``There were some movies that I did not want to watch or movies similar to them. So, I could simply select them on the plot and drag them out [using VbD]. So, I did not have to look at the panel in that case. The panel [MVS] was also useful, when I knew for example that I wanted to filter the movies with specific IMDB Rating values. So it was giving me that accuracy that I needed in that case. So, I think having both together is useful.''} 
Another participant (P10) talked about the benefits of being able to switch between the paradigms on demand: \textit{``Combining two [paradigms] helps a lot with giving a lot of user control. Like I could do it whatever way I prefer to do it. So, if doing one thing in a specific way [paradigm] is not super natural to me then I can do it another way.''}

Several participants \textbf{combined the two interaction paradigms} to perform a single task. 
For instance, participants combined VbD and MVS 12 times to perform \opFilter. 
Many times, they first filtered out a subset of data points by dragging and dropping them onto the Filter Panel (VbD). 
In response, the system recommended different filtering options. 
After accepting a recommendation, participants continued their operation by using the range slider to fine tune their filtering criteria (MVS). 
For example, P2 said during the interview: \textit{``I used two techniques [paradigms] for filtering. Because the demonstration filtering is intuitive but not very precise. So, for precision I fine tuned it using slider.''} P9 also mentioned: \textit{``I prefer do it by demonstration but it is not always very accurate so I had to use sliders on this panel [Filter panel] to get the exact values.''}
In another example, participants first used MVS to map a data attribute to color encoding (\opMark), 
by dragging and dropping the data attribute to the color shelf. 
Then they colored a few data points using VbD to indicate their interest in customizing the color palette. 
The system then recommended color mappings containing the specified colors.

We also noticed that participants sometimes \textbf{switched between paradigms} because they found one paradigm less effective for a given operation. 
For example, three participants switched from VbD to MVS to perform \opMark. 
When we asked participants to explain why they switched to MVS, they mentioned that the system did not recommend what they expected. 
For example, P2 said: \textit{`` [...] when I wanted to assign color to directors, doing it by demonstration I did not get a recommendation that I wanted for that. So I then said lets do it using another technique (MVS).''} 
P3 also noted: \textit{``for cases that suggestions were not accurate I preferred using drag and drop [MVS].''} 

\subsubsection{Which interaction paradigm do people use more often?}
\begin{figure} [t]
\centering
  \includegraphics[width=\columnwidth]{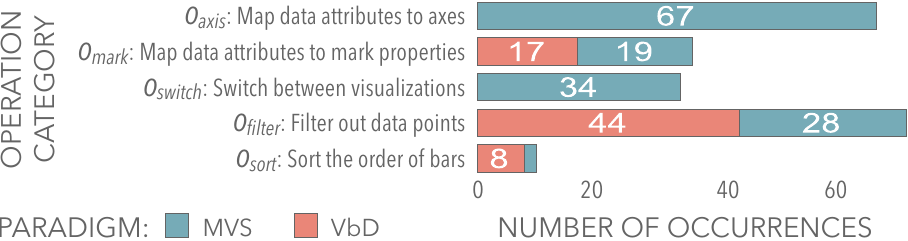}
  \vspace{-1.5em}
\caption{The number of times participants performed each operation using each paradigms.}
\label{fig:operationCount} 
\vspace{-1em}
\end{figure}

Figure~\ref{fig:operationCount} shows the number of times participants performed each operation according to the paradigm they used. 
Participants performed 150 operations using MVS and 69 using VbD in total -- including \opAxis and \opSwitch, the operations that only MVS supports . 
Participants found MVS effective for performing \opAxis and \opSwitch and particularly liked how MVS is consistent for mapping data attributes to different visual encodings.
For all operations that both MVS and VbD support (\opMark, \opFilter, \opSort), \textbf{participants used both paradigms in similar proportions} (50 occurrences using MVS and 69 using VbD).

During our interviews, we asked participants to explain the situations where they found one interaction paradigm more effective/useful than another one. 
8/10 participants found MVS easy to learn and effective for most operations. 
For example, P4 said: \textit{``I used the dragging [MVS] in many cases because I knew what exactly what the outcome would be.''} 
P6 also stated: \textit{``I preferred using drag and drop over demonstration for many tasks because it was easier to use.''} 

Most participants (9/10) found VbD more effective/useful than MVS for \opFilter, for several reasons. 
P2 found performing \opFilter with VbD very intuitive: 
\textit{``I definitely like the filtering by demonstration. Being able to highlight parts and drag them out was I think very intuitive.''}
P9 found the interaction more user-friendly than with MVS: 
\textit{``I use Tableau to analyze data in my research. I don't know if filtering has been this user friendly in Tableau.''}
P7 liked how VbD makes it easier to filter attributes with large cardinalities such as movie directors: 
\textit{``The drag and drop road [MVS] for filtering attributes that have a large number of categorical variables [...] is really hard. So in that case the demonstration road is much easier. I can select instances of the movies that I want to remove on the visualization and drag them out and the system recommends me options for filtering out.''} 

Participants had different takes on which paradigm to use for \opMark. 
P3, P5, and P7 mentioned they prefer using MVS for mapping data attributes to size and color. 
For example, P3 said \textit{``I don't like much about the coloring or resizing using demonstration approach, I would rather just drag the variable to the color using drag and drop [MVS]''} and P5 said \textit{``I feel dragging variables to color is easier than using demonstration'' (P5)}. 
On the other hand, P8 and P9 preferred using VbD for this operation. 
For example, P8 said \textit{``I love this feature because it allows me to color them directly instead of me thinking which attribute should I assign to color, so that these points have the same color. [...] the system then takes attributes that are similar between those points and suggest me attributes [...] That is the beauty of this system.''}
P2 had a unique take on \opMark, stating that his choice of paradigm for this operation relied on the size of the dataset: 
\textit{``it depends on how much data we are dealing. With a small dataset I would prefer doing the coloring by demonstration, but with the larger dataset, I would probably say no let's do drag and drop.''} 

P3, P5, P7, P9 and P10 used VbD to sort bars (\opSort), and only P8 used MVS to sort bars (twice).
Only P10 commented on this operation, saying: \textit{``dragging the bars to sort the bar chart is good. In general it feels very natural to do.''} (VbD).


\section{Discussion}
Results from our qualitative study highlight challenges, possible solutions, and future directions for multi-paradigm interfaces.
Specifically, we discuss the benefits and drawbacks of being able to choose among interaction paradigms, the cost associated with switching between paradigms, the issue of ambiguity in user inputs, and the challenges of learnability and discoverability of such visualization tools. 
Further, we discuss limitations and avenues for future work informed by our experience designing and studying a multi-paradigm tool.

\subsection{Choosing among Interaction Paradigms}
All participants in our study did use multiple paradigms and found the tool empowering and effective.
However, because multi-paradigm tools require people to choose among paradigms while forming a goal, they are likely to increase users' \textbf{cognitive load}.
We found several instances of participants pausing for a few seconds to decide which paradigm to choose for performing the operation at hand. 
For example, while mapping a data attribute to color, P1 said: \textit{``Let me see. hmm. I will go with drag and drop.''} 
In another example, before filtering out some data points, P6 stated: \textit{``not sure which approach to use here.''} These observations echo previous research indicating that even though human intuition may want freedom of choice and flexibility, decisions among choices require mental and visual concentration~\cite{schwartz2004paradox,Ragan2011Effects}.

More generally, multi-paradigm tools might require users to put more effort into thinking about the data and the task at hand, and selecting an interaction paradigm to perform that task. 
This is likely to make the process of visualization construction \textbf{slower} in such tools. 
However, a visualization is not just a means to an end. 
Reflection on the data, tasks at hand, and possible interactions also take place during the data exploration process~\cite{hinrichs2017defense}. 
Optimizing for efficiency may not lead to the overall best outcome, as it may result in users glossing over important details of the data, the operations, and how to best perform the operations. 
Indeed, ``slow'' data exploration can results in users' active involvement, foster creativity and critical thinking, and encourage conscious, deliberate, analytical reasoning~\cite{perin:2014:bertifier,Nissen:2015} -- in other words, System 2 thinking~\cite{kahneman2011thinking}. 
We hypothesize that multi-paradigm tools can bolster deliberate and more logical processes by requiring users to think carefully about their tasks at hand and the different ways of achieving these tasks -- but such hypothesis remains to be studied.

\subsection{Interaction Cost in Switching Between Paradigms}
Switching between paradigms likely increases the \textbf{interaction cost}~\cite{lam:2008:framework} in multi-paradigm tools. 
Users form habits and learn the design principles that went into a particular paradigm when they use that paradigm over a certain period of time~\cite{norman2013design}. 
When switching between paradigms, users might need to recall the principles of each paradigm many times. 
This point is supported by our study, in which participants sometimes felt confused about the functionality of a paradigm after switching to it.
For instance, after P5 had used MVS to drag and drop data attributes to the x and y axes shelves, he said \textit{``let me color using demonstration''}, then selected a subset of data points and dragged them to the color shelf on the encoding panel. 
He then immediately said: \textit{``Oh. I can't do that.''} 
This example illustrates how after forming a habit with one paradigm, switching to another paradigm requires breaking this habit and spending some time to recall the principles of the new paradigm. 

Decreasing the interaction cost that results from switching between paradigms is one of the main challenges for multi-paradigm visualization tools. 
One solution is to design mechanisms such as feedforward~\cite{norman2013design,VermeulenFeedforward} to communicate what can be done with the paradigm that is being used, as well as feedback~\cite{norman2013design} to signify to users that they recall design principles of the other one paradigm.


\subsection{Teaching Multi-paradigm Interfaces in Context}
Combining multiple paradigms into a unified tool can increase functionality supported by the tool. 
This raises new challenges, including that of learning and discovering which functionality is supported by which paradigm. 
Going forward, we envision \textbf{learnability} and \textbf{discoverabiliy} to be a challenge for multi-paradigm tools.

One way to overcome this challenge is to design and include mechanisms to teach multi-paradigm interfaces in context. 
When a user performs an operation using a paradigm, the system can teach her how to perform that operation using another paradigm. 
For example, upon using VbD to map an attribute to color, the system could present how the same operation can be achieved using MVS. This could be achieved by showing an animation where the mapped attribute moves to the color shelf on the Encoding Panel.
The design space of techniques for teaching multi-paradigms in context appears to be large.
Exploring this design space and assessing the effectiveness of such techniques is a promising avenue for future research.

The more people use a tool, the more they learn and discover about the tool~\cite{norman:book93}.
Thus, one could argue that people would discover the features supported by each paradigm as they use the tool over time. 
Moreover, there is a body of previous work providing design guidelines on how to enhance discoverability and learnability in multimodal interfaces~\cite{Srinivasan:2019,turk2014multimodal}. 
We believe multi-paradigm and multi-modal interfaces share similarities, thus investigating the extent to which these guidelines apply to multi-paradigm interfaces looks promising. 
For example, we could envision displaying contextually relevant interaction options on the interface as users interact with a system~\cite{Srinivasan:2019}. This could be achieved by providing recommendations on the interface to show what are the available interactions for the next steps of data exploration.

\subsection{Challenges in Combining Multiple Paradigms in \tool}
Our aim during the design process of \tool was to follow the design goals we introduced in Section 6.1 (\ref{dg:1}--\ref{dg:3}). However, we encountered several challenges following these design goals.

From our preliminary study we learned that not every operation must be supported using both paradigms. However, the main challenge here was to \textbf{identify a set of operations that need to be supported using each as well as both paradigms} (\ref{dg:1}). To design \tool, we mainly relied on the results of our preliminary study and the previous work that measured the effectiveness of different operations using the MVS and VbD paradigms~\cite{saket2018evaluation}. However, going forward we envision studies to measure the effectiveness of different operations for each paradigm prior to implementing a multi-paradigm system.

Another challenge that we encountered when trying to enable paradigms to work in conjunction (\ref{dg:2}) was to take into account screen real estate and keep the interface less occupied with interface elements such as menus and shelves. We approached this challenge by designing \textbf{interface elements that are shared between paradigms}. For example, the Filter Panel is shared between both paradigms in \tool. Users could drag-and-drop data attributes (using MVS) or data points (using VbD) onto the Filter Panel to complete a filtering operation. Sharing interface elements not only enables us to better use the screen real estate, but also enables us to move toward the ultimate goal of \textit{blending} paradigms rather than just supporting two independent paradigms.

Another challenge that we faced when facilitating synchronization between paradigms (\ref{dg:3}) was to understand how we can incorporate changes made by one paradigm into subsequent actions with another paradigm. 
To address this challenge, we \textbf{transferred knowledge between paradigms} by leveraging user interaction with one paradigm to another. For example, if users created a customized color mapping using VbD, the next time the user assigns the same attribute to the color encoding the system preserves the color palette specified by the user.


\subsection{Limitations and Future work}
To investigate the challenge of blending interaction paradigms in a visualization tool, we selected two specific interface designs for the MVS and VbD paradigms. However, user interface design in visualization tools embodying a specific paradigm can be implemented in various ways~\cite{grammel2013survey}. For instance, a tool that embodies MVS could be implemented using the shelf configuration, data flow, or visual builder interface design. Each of these implementation variations would influence the construction process differently. 
As such, we emphasize that the selected designs in this study do not represent all possible interface designs for the MVS and VbD paradigms, and that our results unlikely apply fully to all possible designs. 
Thus, we encourage future work to consider the effect of tool design when building on our findings.

Finally, we did not control for participants' expertise. 
Expertise and prior knowledge about visualizations likely influence participants' visualization construction process. 
For instance, expert users might prefer using the MVS paradigm since they have a better understanding of visual encodings, or because they are familiar with existing tools that leverage MVS. 
In contrast, novice users might use VbD more often because of the freedom of expression it offers, and because it does not require users to formalize the mappings between the data and visual encodings. 
However, this remains to be formally studied. 
Our work can be extended with other studies such as investigating multi-paradigm interfaces for users with different levels of visualization expertise.

\section{Conclusions}
We investigated how to offer the benefits of two different interaction paradigms (Manual View Specification and Visualization by Demonstration) that use the same input modality (mouse) in a unified visualization tool. 
We first presented \tool, a testbed to investigate opportunities and challenges in combining multiple paradigms in a single tool. 
We then conducted a study to understand how people use both interaction paradigms for visualization construction and data exploration. 

Our findings provide evidence that people 
1) use both paradigms interchangeably, 
2) seamlessly switch between paradigms based on the operation at hand, 
and 3) choose to combine paradigms to successfully complete a single operation.
These findings provide the first list of empirically-grounded benefits and challenges of combining interaction paradigms in visualization tools, 
paving the way for promising future research directions -- 
including decreasing the interaction cost that results from switching between paradigms, 
addressing ambiguity in user inputs,
and teaching multi-paradigm interfaces in context.


\bibliographystyle{abbrv-doi}

\bibliography{template}
\end{document}